\documentclass[aps,preprint,floats,epsf,epsfig,nofootinbib,letter]{revtex4}

\usepackage{graphicx}
\usepackage{dcolumn}
\usepackage{bm}

\def\be{\begin{eqnarray}}
\def\en{\end{eqnarray}}
\def\non{\nonumber}

\def\CP{{\it CP}~}
\def\cp{{\it CP}}

\begin{document}

\renewcommand{\baselinestretch}{1.10}

\font\el=cmbx10 scaled \magstep2{\obeylines\hfill May, 2011}

\vskip 1.5 cm

\centerline{\Large\bf Remarks on the Qin-Ma Parametrization }
\centerline{\Large\bf of Quark Mixing Matrix}

\bigskip
\centerline{\bf Y. H. Ahn\footnote{Email: yhahn@phys.sinica.edu.tw},
Hai-Yang Cheng\footnote{Email: phcheng@phys.sinica.edu.tw},
and Sechul Oh\footnote{Email: scoh@phys.sinica.edu.tw}}
\medskip
\centerline{Institute of Physics, Academia Sinica}
\centerline{Taipei, Taiwan 115, Republic of China}
\medskip

\medskip

\centerline{\bf Abstract}
\bigskip
\small
Recently, Qin and Ma (QM) have advocated a new Wolfenstein-like parametrization of the quark mixing
matrix based on the triminimal expansion of the Cabibbo-Kobayashi-Maskawa (CKM) parametrization. The
\cp-odd phase in the QM parametrization is around $90^\circ$ just as that in the CKM parametrization.
We point out that the QM parametrization can be readily obtained from the Wolfenstein parametrization
after appropriate phase redefinition for quark fields and that the phase $\delta$ in both QM and CKM
parametrizations is related to the unitarity angles $\alpha$, $\beta$ and $\gamma$, namely,
$\delta= \beta+\gamma$ or $\pi-\alpha$.
We show that both QM  and Wolfenstein parametrizations can be deduced from the CKM and Chau-Keung-Maiani
ones. By deriving the QM parametrization from the exact Fritzsch-Xing (FX) parametrization of the quark mixing
matrix, we find that the phase of the FX form is in the vicinity of $-270^\circ$ and hence
$\sin\delta\approx 1$. We discuss the seeming discrepancy between the Wolfenstein and QM parametrizations
at the high order of $\lambda\approx |V_{us}|$.

\maketitle

%
%

{\bf 1.}  In the standard model with three generations of quarks, the $3\times 3$ unitary quark mixing
matrix
\be
V=\left( \matrix{V_{ud} & V_{us} & V_{ub}  \cr   V_{cd} & V_{cs} &
V_{cb}   \cr  V_{td} & V_{ts} & V_{tb}  \cr}\right)
\en
can be parametrized in infinitely many ways with three rotation angles and one \cp-odd phase.
All different parametrizations lead to the same physics. A well-known simple parametrization introduced
by  Wolfenstein \cite{Wolfenstein:1983yz} is
\be \label{eq:Wolf}
V_{{\rm Wolf}}=\left( \matrix{ 1-\lambda^2/2 & \lambda &
A\lambda^3(\rho-i\eta)   \cr   -\lambda & 1-\lambda^2/2   & A\lambda^2   \cr
A\lambda^3(1-\rho-i\eta) & -A\lambda^2   & 1  \cr}\right)+{\cal O}(\lambda^4).
\en
Using the global fits to the data, the four unknown real parameters $A$, $\lambda$, $\rho$ and $\eta$
are determined to be
\be
A=0.812^{+0.013}_{-0.027}\,, \quad \lambda=0.22543\pm 0.00077\,, \quad \bar\rho=0.144\pm0.025\,,
\quad \bar\eta=0.342^{+0.016}_{-0.015}\,,
\en
by the CKMfitter Collaboration \cite{CKMfitter} and
\be
A=0.807\pm0.01\,, \quad \lambda=0.22545\pm 0.00065\,, \quad \bar\rho=0.143\pm0.03\,,
\quad \bar\eta=0.342\pm0.015\,,
\en
by the UTfit Collaboration \cite{UTfit}, where $\bar \rho=\rho(1-\lambda^2/2+\cdots)$ and
$\bar\eta=\eta(1-\lambda^2/2+\cdots)$.

Recently, Qin and Ma (QM)~\cite{Qin:2010hn} have advocated a new Wolfenstein-like
parametrization of the quark mixing matrix
\be \label{eq:QM}
V_{{\rm QM}}=\left( \matrix{ 1-\lambda^2/2 & \lambda & h\lambda^3 e^{-i\delta_{\rm QM}}  \cr
-\lambda & 1-\lambda^2/2   & (f+h e^{-i\delta_{\rm QM}})\lambda^2   \cr
f\lambda^3 & -(f+h e^{i\delta_{\rm QM}})\lambda^2   & 1  \cr}\right)+{\cal O}(\lambda^4),
\en
based on the triminimal expansion of the Cabibbo-Kobayashi-Maskawa (CKM) matrix.
It is obvious that once the parameters $f$ and $h$ are fixed from the matrix elements $V_{td}$ and
$V_{ub}$, respectively, the phase $\delta_{\rm QM}$ is ready to be determined from the measurement of
$V_{cb}$.
From the global fits to the quark mixing matrix given below in Eq. (\ref{eq:Vfit}) we obtain
\be \label{eq:QMfh}
f=0.749^{+0.034}_{-0.037}\,,\qquad h=0.309^{+0.017}_{-0.012}\,,
 \qquad \delta_{\rm QM}=(89.6^{+2.94}_{-0.86})^\circ\,.
\en
Therefore, \CP violation is approximately maximal in the sense that $\sin\delta\approx 1$. Indeed, it is known
that  the phase in the Kobayashi-Maskawa parametrization is also in the vicinity of maximal \CP
violation. We shall show below that by rephasing the Wolfenstein parametrization, it is easily seen
why the phase $\delta_{\rm QM}$ of the Qin-Ma parametrization and $\delta_{\rm KM}$ of the
Kobayashi-Maskawa one are both of order $90^\circ$.

Since the phases of the matrix elements $V_{ub}$ and $V_{td}$ in the Wolfenstein parametrization are
$\arctan(\eta/\rho)\approx \gamma$ and $\arctan(\eta/(1-\rho))\approx \beta$, respectively, it has been
argued in \cite{Qin:2010hn} that ``one has difficulty to arrive at the Wolfenstein parametrization from
the triminimal parametrization of the KM matrix". The purpose of this short note is to point out that
both Wolfenstein and Qin-Ma parametrizations can be obtained easily from the Cabibbo-Kobayashi-Maskawa
and Chau-Keung-Maiani matrices to be discussed below.  Koide \cite{Koide:2004gj} pointed out that
among the possible parametrizations of the quark mixing matrix, only the CKM and the Fritzsch-Xing
(FX)~\cite{Fritzsch:1997fw} parametrizations can allow to have maximal \CP violation. In this work, we
are going to show that the QM parametrization derived from the FX parametrization will enable us to
see the feature of maximal \CP nonconservation in the FX form.  We shall also compare the Wolfenstein
and QM parametrizations at the high order of $\lambda$.

\vskip 0.3cm {\bf 2.}
The well-known Cabibbo-Kobayashi-Maskawa parametrization of $V$ is given by \cite{Kobayashi:1973fv}
\be
V_{\rm CKM} =\left( \matrix{c_1 &
-s_1c_3  & -s_1s_3  \cr
s_1c_2 & c_1c_2c_3-s_2s_3e^{i\delta_{\rm KM}} & c_1c_2s_3+s_2c_3
e^{i\delta_{\rm KM}} \cr  s_1s_2 & c_1s_2c_3+c_2s_3e^{i\delta_{\rm KM}} & c_1s_2s_3-c_2c_3
e^{i\delta_{\rm KM}}  \cr}\right),
\en
where $c_i\equiv\cos\theta_i$ and $s_i\equiv\sin\theta_i$. Using the matrix elements of $|V|$ determined
from global fits at 1$\sigma$ level \cite{CKMfitter}
\be \label{eq:Vfit}
\left( \matrix{0.97425\pm 0.00018 & 0.22543^{+0.00077}_{-0.00077} &0.00354^{+0.00016}_{-0.00014} \cr
0.22529\pm0.00077  & 0.97342^{+0.00021}_{-0.00019} & 0.04128^{+0.00058}_{-0.00129} \cr
0.00858^{+0.00030}_{-0.00034}  & 0.04054^{+0.00057}_{-0.00129} &
0.999141^{+0.000053}_{-0.000024} \cr}\right),
\en
we obtain
\be \label{eq:CKMangles}
 \theta_1 = ( 13.03 \pm 0.05 )^{\circ},
 \quad \theta_2 = ( 2.18^{+0.08}_{-0.09} )^{\circ} \quad
 \theta_3 = ( 0.90^{+0.044}_{-0.039} )^{\circ},\quad \delta_{\rm KM}=(88.88^{+4.11}_{-2.05})^\circ.
\en
There is one disadvantage in this parametrization, namely, the matrix element $V_{tb}$ has a large
imaginary part. Since \cp-violating effects are known to be  small, it is thus desirable to parameterize
the mixing matrix in such a way that the imaginary part appears with a smaller coefficient.
The parametrization proposed by Maiani in 1977 \cite{Maiani}
\be \label{eq:Maiani}
V_{{\rm Maiani}}=\left( \matrix{ c_{12}c_{13} & s_{12}c_{13}  & s_{13} \cr
  -s_{12}c_{23}-c_{12}s_{23}s_{13} e^{i\phi}  & c_{12}c_{23}-s_{12}s_{23}s_{13}e^{i\phi} &
  s_{23}c_{13}e^{i\phi}   \cr
  s_{12}s_{23}e^{-i\phi}-c_{12}c_{23}s_{13}
& -c_{12}s_{23}e^{-i\phi}-s_{12}c_{23}s_{13} & c_{23}c_{13}   \cr}\right)
\en
has the nice feature that its imaginary part is proportional
to $s_{23}\sin\phi$, which is of order $10^{-2}$.
In 1984 Chau and Keung introduced another parametrization \cite{Chau:1984fp} (see also \cite{Harari:1986xf})
\be \label{eq:CK}
V_{\rm CK}=V_{\overline{\rm CKM}}=\left( \matrix{ c_{12}c_{13} & s_{12}c_{13}  & s_{13}e^{-i\phi} \cr
-s_{12}c_{23}-c_{12}s_{23}s_{13}
e^{i\phi}  & c_{12}c_{23}-s_{12}s_{23}s_{13}e^{i\phi} &   s_{23}c_{13}   \cr  s_{12}s_{23}-c_{12}c_{23}s_{13}
e^{i\phi}   & -c_{12}s_{23}-s_{12}c_{23}s_{13}e^{i\phi} & c_{23}c_{13}   \cr}\right),
\en
which is equivalent to the Maiani parametrization after the quark field redefinition: $t\to t\, e^{i\phi}$
and $b\to b\, e^{-i\phi}$. It is evident that the imaginary part in this parametrization is proportional
to $s_{13}\sin\phi$, of order $10^{-3}$.
This Chau-Keung-Maiani (another CKM !) parametrization  denoted by $V_{\overline{\rm CKM}}$ or $V_{\rm CK}$
has been advocated by the Particle Data Group (PDG) \cite{PDG10} to be
the standard parametrization for the quark mixing matrix.
\footnote{The Maiani parametrization Eq. (\ref{eq:Maiani}) was once proposed by PDG (1986 edition) \cite{PDG86}
to be the standard parametrization for the quark mixing matrix.}
It follows from Eqs. (\ref{eq:Vfit}) and (\ref{eq:CK}) that
\begin{eqnarray} \label{eq:CKqmix}
  \theta_{12} = ( 13.03 \pm 0.05 )^{\circ},
  ~~\theta_{23} = ( 2.37^{+0.03}_{-0.07} )^{\circ}, ~
  ~~\theta_{13} = ( 0.20^{+0.01}_{-0.01} )^{\circ},
  ~~\phi = ( 67.19^{+2.40}_{-1.76} )^{\circ} ~.
 \end{eqnarray}

The Wolfenstein parametrization can be easily obtained from the  exact $\overline{\rm CKM}$ parametrization by
using the relations
\be \label{eq:CKtoW}
s_{12}= \lambda, \qquad s_{23}=A\lambda^2, \qquad s_{13}e^{-i\phi}=A\lambda^3(\rho-i\eta).
\en
It should
be stressed that the Wolfenstein parametrization given in Eq. (\ref{eq:Wolf}) is just an approximation
to order $\lambda^3$ and sometimes it may give a wrong result if higher order $\lambda$ terms are not
included. For example, to ${\cal O}(\lambda^3)$ the rephasing-invariant quantity
$F=V_{ud}V_{cs}V_{us}^*V_{cd}^*$ is real and hence cannot induce \CP violation. In order to obtain the
imaginary part of $F$, one has to expand Eq. (\ref{eq:Wolf}) to the accuracy of ${\cal O}(\lambda^5)$
(see Eq. (\ref{eq:Vhighorder}) below). To derive the Wolfenstein parametrization from the CKM one,
we first rotate the phases of some of the quark fields
$s\to s\,e^{i\pi}$, $c\to c\,e^{i\pi}$, $b\to b\,e^{-i(\theta+\pi)}$,
$t\to t\,e^{-i(\delta_{\rm KM}-\theta)}$
and substitute the relations
\be \label{eq:KMtoW}
s_1= \lambda, \qquad s_2e^{-i(\delta_{\rm KM}-\theta)}=A\lambda^2(1-\rho-i\eta), \qquad
s_{3}e^{-i\theta}=A\lambda^2(\rho-i\eta)
\en
in the CKM parametrization to obtain the Wolfenstein one. From the above equation we are led to
\be
\delta_{\rm KM}=\arctan \left({\eta\over \rho}\right)
  +\arctan\left({\eta\over 1-\rho}\right) \approx \gamma+\beta=\pi-\alpha,
\en
where the three angles $\alpha$, $\beta$ and $\gamma$ of the unitarity triangle
are defined by
\be
\alpha\equiv{\rm arg}\left(-{V_{td}V^*_{tb}\over V_{ud}V_{ub}^*}\right),~~~
\beta\equiv{\rm arg}\left(-{V_{cd}V^*_{cb}\over V_{td}V_{tb}^*}\right),~~~
\gamma\equiv{\rm arg}\left(-{V_{ud}V^*_{ub}\over V_{cd}V_{cb}^*}\right),
\en
and they satisfy the relation $\alpha+\beta+\gamma=\pi$.
Since $\alpha=(91.0\pm3.9)^\circ$, $\beta=(21.76^{+0.92}_{-0.82})^\circ$ and $\gamma=(67.2\pm3.0)^\circ$ \cite{CKMfitter},
the phase $\delta_{\rm KM}$ is thus very close to $90^\circ$.

It is also easily seen that the phase $\delta_{\rm QM}$ of the Qin-Ma parametrization can be
expressed in terms of the unitarity angles $\alpha, ~\beta$ and $\gamma$.
Starting from the Wolfenstein parametrization in Eq.~(\ref{eq:Wolf}) with
$V_{ub}= |V_{ub}| e^{-i \gamma}$ and $V_{td}=|V_{td}| e^{-i \beta}$, one can rephase
the $t$ and $b$ quark fields as
 $t \to t \,e^{i \beta}$, $b \to b \,e^{-i \beta}$.
Substituting the relations
\be \label{eq:A}
A\sqrt{(1-\rho)^2+\eta^2}=f, \qquad A\sqrt{\rho^2+\eta^2}=h, \qquad
Ae^{-i\beta}=f+he^{-i\delta_{\rm QM}},
\en
in the Wolfenstein parametrization,
we see that the QM parametrization is obtained  with  the phases
of $V_{ub}$ and $V_{cb}$ expressed as
\be
 \delta_{\rm QM} = \gamma +\beta = \pi -\alpha ~, \qquad
 {\rm arctan}\Big( \frac{h \sin\delta_{\rm QM}}{f +h \cos\delta_{\rm QM}} \Big) = \beta ~.
\en
Therefore, the phases $\delta_{\rm QM}$ and $\delta_{\rm KM}$ are both equal to $\gamma+\beta$ or
$\pi-\alpha$. Note that Eq. (\ref{eq:A}) leads to $A=(f^2+h^2)^{1/2}$ and
$A(1-2\rho+2\rho^2+2\eta^2)^{1/2}=(f^2+h^2)^{1/2}$. These two relations are consistent with each other
as $(1-2\rho+2\rho^2+2\eta^2)^{1\over 2}=0.9993$ is very close to 1\,.

It is straightforward to obtain the Qin-Ma parametrization from the CKM matrix by making the phase rotation
$s\to s\,e^{i\pi}$, $c\to c\,e^{i\pi}$, $b\to b\,e^{i(\pi-\delta_{\rm KM})}$,
followed by the replacement
\be \label{eq:KMtoQM}
s_1= \lambda, \qquad s_2=f\lambda^2, \qquad s_3e^{-i\delta_{\rm KM}}=h\lambda^2e^{-i\delta_{\rm QM}}.
\en
As a result, $\delta_{\rm QM}=\delta_{\rm KM}$,
as it should be.
For the  $\overline{\rm CKM}$ matrix, the Qin-Ma parametrization is obtained by first performing the quark
field redefinition $b\to b\,e^{-i\theta}$ and $t\to t\,e^{i\theta}$ and then adapting the relations
\be \label{eq:CKtoQM}
s_{12}= \lambda, \qquad s_{23}e^{-i\theta}=(f+he^{-i\delta_{\rm QM}})\lambda^2, \qquad
  s_{13}e^{-i(\phi+\theta)}=h\lambda^3e^{-i\delta_{\rm QM}}.
\en

From Eqs. (\ref{eq:CKtoW}), (\ref{eq:KMtoW}), (\ref{eq:KMtoQM}) and (\ref{eq:CKtoQM}) we see that
\be
&& {s_{13}\over s_{23}} =\sqrt{\rho^2+\eta^2}\,\lambda={h\over \sqrt{f^2+h^2}}\lambda =0.38\lambda\,, \non \\
&& {s_3\over s_2} ={h\over f}={\sqrt{\rho^2+\eta^2}\over \sqrt{(1-\rho)^2+\eta^2}} =0.41\,,
\en
and
\be
s_1:s_2: s_3 = 1: 0.75\lambda : 0.31\lambda\,, \qquad
s_{12}:s_{23}: s_{13} = 1: 0.81\lambda : 0.31\lambda^2.
\en
Therefore, the hierarchial pattern for the three mixing angles in the Cabibbo-Kobayashi-Maskawa and
Chau-Keung-Maiani parametrizations is very similar for the first two angles but different for the
third angle. The corresponding Jarlskog invariant $J$ \cite{Jarlskog:1985ht} has the expression
\footnote{The concept of rephasing invariance for physical quantities and the use of the rephasing
invariant quantity $J$ became popular in the early and middle eighties.
Historically,  Chau and Keung already pointed out in their 1984 seminal paper that all \cp-violating
effects  are proportional to a universal factor
which they called $X_{CP}$ \cite{Chau:1984fp}. They showed explicitly that the quantity
Im[$V_{ij}V_{kl}V_{il}^*V_{kj}^*$] is proportional to $X_{CP}$.
}
\be
J_{\rm CKM} &\approx& s_1^2s_2s_3\sin\delta_{\rm KM}= fh\lambda^6, \non \\
J_{\overline{\rm CKM}} &\approx&
s_{12}s_{23}s_{13}\sin\phi=A^2\eta\lambda^6,
\en
with the magnitude of $3.0\times 10^{-5}$.

\vskip 0.3cm {\bf 3.}
Fritzsch and Xing \cite{Fritzsch:1997st} were the first (see also \cite{Rasin:1997pn}) to point out
that there exist nine fundamentally different  ways to describe the quark mixing matrix.
\footnote{Of course, the freedom of rotating the phase of quark fields will render the parametrization
of the quark mixing matrix infinitely many.}
Moreover, they argued that the Fritzsch-Xing (FX) parametrization proposed by them \cite{Fritzsch:1997fw}
\be
V_{{\rm FX}}=\left( \matrix{ s_xs_yc_z+c_xc_ye^{-i\phi_{\rm FX}} & c_xs_yc_z-s_xc_ye^{-i\phi_{\rm FX}} &
  s_ys_z \cr
  s_xc_yc_z-c_xs_ye^{-i\phi_{\rm FX}} & c_xc_yc_z+s_xs_ye^{-i\phi_{\rm FX}}  & c_ys_z \cr
  -s_xs_z   & -c_xs_z & c_z   \cr}\right),
\en
in which the \cp-violating phase resides solely in the light quark sector, stands up as the most favorable
description of the flavor mixing. As shown in \cite{Koide:2004gj}, among the nine distinct parametrizations,
only the CKM and FX parametrizations allow to have maximal \CP violation. To see this is indeed the case
for the FX form, let us derive the QM parametrization from it.

Substituting the relations
\be \label{eq:FXtoQM}
s_x={f\over \sqrt{f^2+h^2}}\,\lambda, \quad
s_y={h\over \sqrt{f^2+h^2}}\,\lambda, \quad
s_z=\sqrt{f^2+h^2}\,\lambda^2,
\en
in the FX parametrization leads to
\be
V_{{\rm FX}}=\left( \matrix{ (1-\lambda^ 2/2) e^{-i\phi_{\rm FX}} & \lambda e^{i\theta_{\rm FX}}  & h\lambda^3 \cr
-\lambda e^{-i(\phi_{\rm FX}+\theta_{\rm FX})} & 1-\lambda^2/2  & \sqrt{f^2+h^2}\,\lambda^2 \cr
  -f\lambda^3 & -\sqrt{f^2+h^2}\,\lambda^2 & 1   \cr}\right)+{\cal O}(\lambda^4),
\en
with
\be
\sin\theta_{_{\rm FX}}={f\over \sqrt{f^2+h^2}}\sin\phi_{_{\rm FX}}, \qquad
\cos\theta_{_{\rm FX}}={h\over \sqrt{f^2+h^2}}\,(1-{f\over h}\cos\phi_{_{\rm FX}}).
\en
Then, making the quark field redefinition
\be
u\to u^{i\phi_{\rm FX}}, \quad c\to c\, e^{i(\phi_{\rm FX}+\theta_{\rm FX})}, \quad t\to t\,e^{-i\pi},
  \quad s\to s\, e^{-i(\phi_{\rm FX}+\theta_{\rm FX})}, \quad b \to b\,e^{i\pi},
\en
and setting
\be
\phi_{_{\rm FX}}=-(\delta_{\rm QM}+\pi),
\en
we finally arrive at the Qin-Ma parametrization (\ref{eq:QM}).
Since $\delta_{\rm QM}$ is around $90^\circ$, it is clear that the phase $\phi_{_{\rm FX}}$ in the
vicinity of $-270^\circ$ leads to maximal \CP violation with $\sin\phi_{_{\rm FX}}=1$.
The corresponding Jarlskog invariant is
\be
J_{_{\rm FX}}=s_xs_ys_z^2\sin\phi_{_{\rm FX}}=fh\lambda^6 \ .
\en

From Eq. (\ref{eq:FXtoQM}) we obtain
\be
s_x:s_y:s_z =1: {h\over f}: {f^2+h^2\over f}\lambda=1:0.41: 0.88\lambda\,.
\en
As a consequence, the hierarchical pattern for the mixing angles in the FX parametrization differs from
that in CKM and $\overline{\rm CKM}$ ones. Recall that the parameter $\lambda$ is equal to $s_1$
($s_{12}$) in the CKM ($\overline{\rm CKM}$) parametrization, while it is identical to $\sqrt{s_x^2+s_y^2}$ in
the FX parametrization.  From Eq. (\ref{eq:Vfit}) we obtain the mixing angles
\be
\theta_x=(11.95^{+0.83}_{-0.64})^\circ, \qquad \theta_y=(4.90^{+0.39}_{-0.26})^\circ, \qquad
\theta_z=(2.38^{+0.07}_{-0.03})^\circ.
\en

\vskip 0.3cm {\bf 4.}
In future experiments such as LHCb and Super $B$ ones, more precise measurements of the CKM matrix
elements are expected so that high order $\lambda$ terms of the CKM matrix elements become more important.
In principle, the expression of the Wolfenstein and QM parametrizations to the high order of $\lambda$ can
be obtained from the exact parametrization of the quark mixing matrix by expanding it to the desired
order of $\lambda$. For example, the substitution of the relations (\ref{eq:CKtoW}) in the
$\overline{\rm CKM}$ matrix for the Wolfenstein parametrization~\cite{Buras:1994ec} and relations
(\ref{eq:KMtoQM}) in the CKM matrix for the QM parametrization~\cite{Qin:2010hn} lead to
\be  \label{eq:Vhighorder}
V_{{\rm Wolf}}&=&
\left( \begin{array}{ccc}
  1 -\frac{\lambda^2}{2} -\frac{\lambda^4}{8} & \lambda & A \lambda^3 (\rho -i \eta)  \nonumber \\
  -\lambda +\frac{1}{2} A^2 \lambda^5 (1 -2\rho -2i \eta) &
    1 -\frac{\lambda^2}{2} -\frac{1}{8} \lambda^4 (1 +4 A^2)
    & A \lambda^2
    \nonumber \\
  A \lambda^3 (1 -\rho -i \eta) +\frac{1}{2} A \lambda^5 (\rho +i \eta)
    & ~ -A \lambda^2 +\frac{1}{2} A \lambda^4 (1 -2\rho -2i \eta) & ~ 1 -\frac{1}{2} A^2 \lambda^4
\end{array} \right)  \nonumber \\
&& +{\cal O}(\lambda^6) ~, \\
V_{{\rm QM}}&=&
\left( \begin{array}{ccc}
  1 -\frac{\lambda^2}{2} -\frac{\lambda^4}{8} & \lambda -\frac{h^2 \lambda^5}{2}
    & h\lambda^3 e^{-i \delta_{\rm QM}}  \nonumber \\
  -\lambda +\frac{f^2 \lambda^5}{2} & ~ 1 -\frac{\lambda^2}{2}
    -\frac{1}{8} (1 +4 f^2 +8 fh e^{i \delta_{\rm QM}} +4 h^2) \lambda^4
    & ~ (f +h e^{-i \delta_{\rm QM}}) \lambda^2 -\frac{1}{2} h \lambda^4 e^{-i \delta_{\rm QM}}
    \nonumber \\
  f \lambda^3 & -(f +h e^{i\delta_{\rm QM}}) \lambda^2 +\frac{1}{2} f \lambda^4
    & 1 -\frac{1}{2} (f^2 +2 fh e^{-i \delta_{\rm QM}} +h^2) \lambda^4
\end{array} \right)  \nonumber \\
&& +{\cal O}(\lambda^6) ~,  \nonumber
\en
up to the order of $\lambda^5$.
However, care must be taken when one compares higher order terms in two different parametrizations.
The point is that when $\lambda$ is treated as an expansion parameter for the quark mixing matrix, the
other parameters should be of order unity. We know this is not the case in reality: the parameters
$h,~\rho$ and $\eta$ are of order $\lambda$ numerically, while $A$, $f$ and $\delta$ are of order unity.
This fact leads to the seeming discrepancy between the corresponding elements of $V_{\rm Wolf}$ and
$V_{\rm QM}$.
For instance, taking into account $h,~\rho$ and $\eta$ being of order $\lambda$ numerically, the order
$\lambda^5$ terms in $V_{us}$ of $V_{\rm QM}$ and in $V_{td}$ of $V_{\rm Wolf}$ are effectively of
order $\lambda^7$ and $\lambda^6$ being negligible, respectively.
Likewise, for $V_{cb}$ of $V_{\rm QM}$, the physical (rephasing-invariant) observable $|V_{cb}|$ is
obtained as
\be \label{Vcb_QM}
|V_{cb}| \approx \lambda^2 \sqrt{f^2 +h^2}~ \left[ 1 - \frac{1}{2} ~\frac{h^2}{f^2 +h^2} ~\lambda^2
  +{\cal O}(\lambda^4) \right] ~,
\en
where $\delta_{\rm QM} \approx 90^{\circ}$ has been used.
With the relation $A \approx \sqrt{f^2 +h^2}$, the correction to the leading term of order $\lambda^2$
starts effectively at order $\lambda^6$ being negligible.
Thus, all the seeming discrepancies between the corresponding elements of $V_{\rm Wolf}$ and $V_{\rm QM}$
are resolved.

\vskip 0.3cm {\bf 5.}
In this work  we have shown
that the Qin-Ma parametrization can be easily obtained from the Wolfenstein parametrization after
appropriate phase redefinition for quark fields and that
the phase $\delta$ in both QM and CKM parametrizations is related to the unitarity angles $\alpha$,
$\beta$ and $\gamma$, namely, $\delta= \beta+\gamma$ or $\pi-\alpha$.
Both QM  and Wolfenstein parametrizations can be deduced from the CKM and Chau-Keung-Maiani
ones. By deriving the QM parametrization from the exact Fritzsch-Xing parametrization, we find that the phase of
the FX form is approximately maximal. From the analysis of this
work, it is easy to see the hierarchial patterns for the quark mixing angles in various different
parametrizations. Finally, we compare the Wolfenstein and QM parametrizations at the high order of
$\lambda$ and point out that all the seeming discrepancies between them are gone when the small parameters
$h$, $\rho$ and $\eta$ are counted as of order $\lambda$.

\vskip 1.4cm {\bf Acknowledgments}

We wish to thank Bo-Qiang Ma for bringing the Qin-Ma parametrization to our attention and for fruitful
discussion. This research was supported in part by the National Science Council of R.O.C. under Grant Nos.
NSC-97-2112-M-008-002-MY3, NSC-97-2112-M-001-004-MY3 and NSC-99-2811-M-001-038.

\end{document}